\documentstyle[11pt,newpasp,twoside]{article}
\def\edcomment#1{\iffalse\marginpar{\raggedright\sl#1\/}\else\relax\fi}
\reversemarginpar

\begin{document}
\title{Processed and unprocessed ices in circumstellar disks}
\author{Klaus Pontoppidan$^1$}
\affil{$^1$Leiden Observatory, P.O.Box 9513, 2300 RA Leiden, Netherlands}
\author{Emmanuel Dartois$^2$, Wing-Fai Thi$^3$, Ewine van Dishoeck$^1$}
\affil{$^2$IAS, B{\^a}t 121, Universit{\'e} Paris XI, 91405 Orsay Cedex, France}
\affil{$^3$Astronomical Institute ``Anton Pannekoek'', University of Amsterdam, Kruislaan 403, 
1098 SJ Amsterdam, Netherlands}

\begin{abstract}
We present 3-5\,$\mu$m VLT-ISAAC spectroscopy searching for evidence of
methanol ices in edge-on disks of young embedded stars. Examples include the disks 
of L1489 IRS in Taurus and CRBR 2422.8-3423 in Ophiuchus, the last of which has
the highest column density of solid CO known toward a YSO. We find no unambiguous evidence
for abundant methanol in the observed disks, but give strict upper limits.
Several additional low-mass sources in the 
Serpens and Chameleon molecular clouds exhibit abundant solid methanol, although it is not 
clear if the ice is associated with a disk or with the envelope. These are the first detections 
of solid methanol in the circumstellar environments of embedded young low-mass stars 
providing evidence that complex molecular species previously observed only in the solid state 
toward high-mass star forming regions are also present near solar-type young stars. The 
constraints on the formation mechanisms of methanol and the chemical evolution of ices as 
the material is incorporated into circumstellar disks are discussed.

\end{abstract}

\section{Introduction}

The presence of abundant ices in the interstellar medium has long been
firmly established and its connection to the ices present in solar-system
comets has been the subject of some discussion. Most of the major constituents
of interstellar grain mantles have now been firmly identified, mostly
due to the ISO mission and extensive laboratory work. However, only recently
have attempts been made to observe the ices in the context of their astrophysical
environments. Models indicate that ices may undergo significant transformation
when material from a protostellar envelope is incorporated into
a disk, either through shock evaporation (Charnley, this volume) and recondensation
or through growth of ice mantles under different densities and temperatures
than those of the quiescent interstellar medium (e.g. Aikawa et al. 2003).

In this contribution, new direct 
observations of ices in circumstellar disks are described and compared to ices more likely
to be associated with circumstellar envelopes and quiescent molecular cloud material. 

\section{Observations of ices in disks}

Mid-infrared spectroscopy of nearly edge-on circumstellar disks around young stars probe 
lines of sight passing through the cold disk mid-planes
where ices are expected to exist. As part of a 3-5\,$\mu$m spectroscopic survey of 
$>$ 40 young low mass stars (van Dishoeck et al. 2003) a small selection of sources with edge-on disks 
have been included.  Previously, the stretching mode of solid CO
at 4.67\,$\mu$m has been analysed for the low mass edge-on disk sources
L1489 IR (Boogert et al. 2002) and CRBR 2422.8-3423 (Thi et al. 2002).

\begin{figure}
\plotfiddle{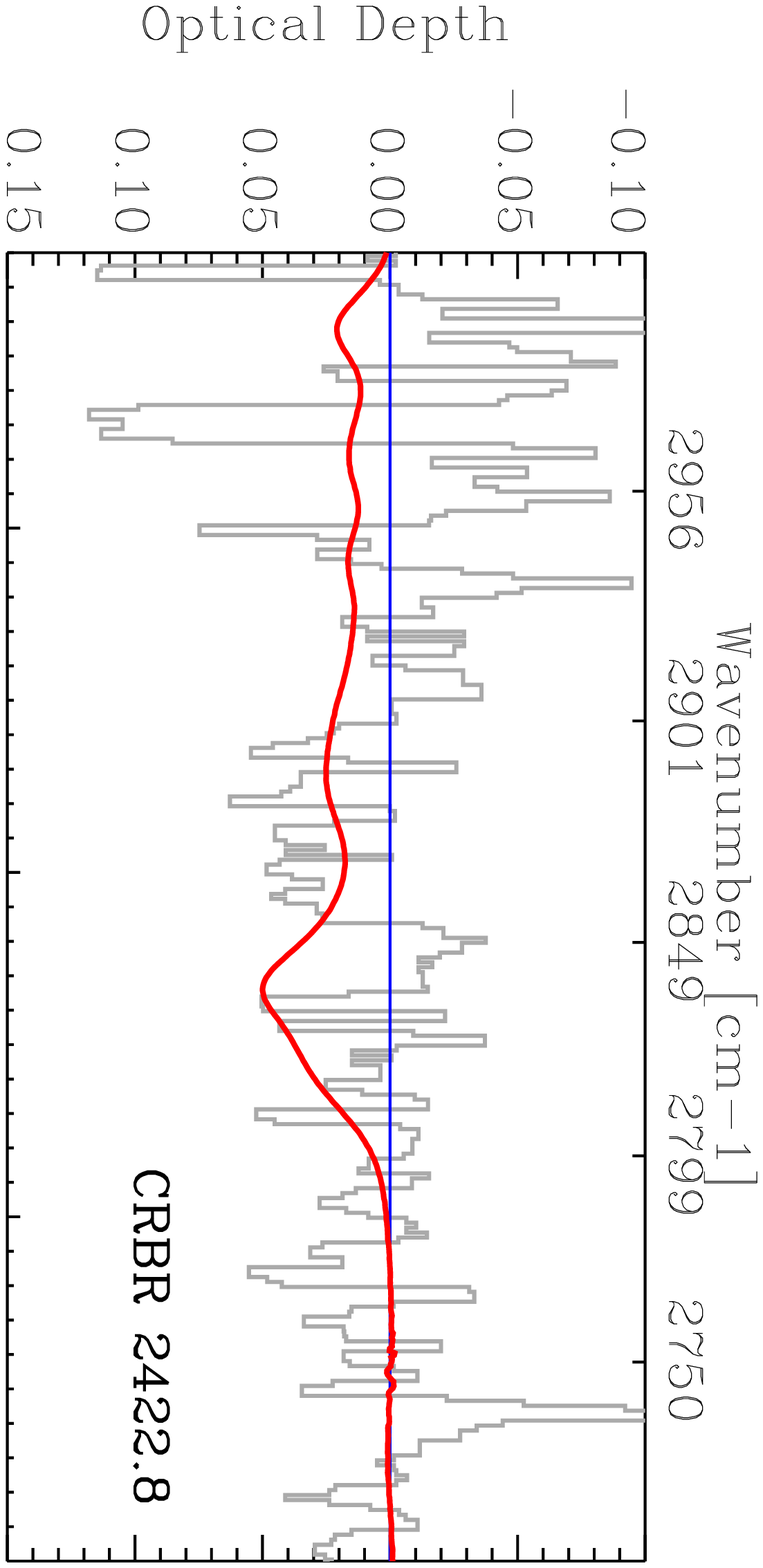}{3.0cm}{90}{30}{30}{100}{-40}
\plotfiddle{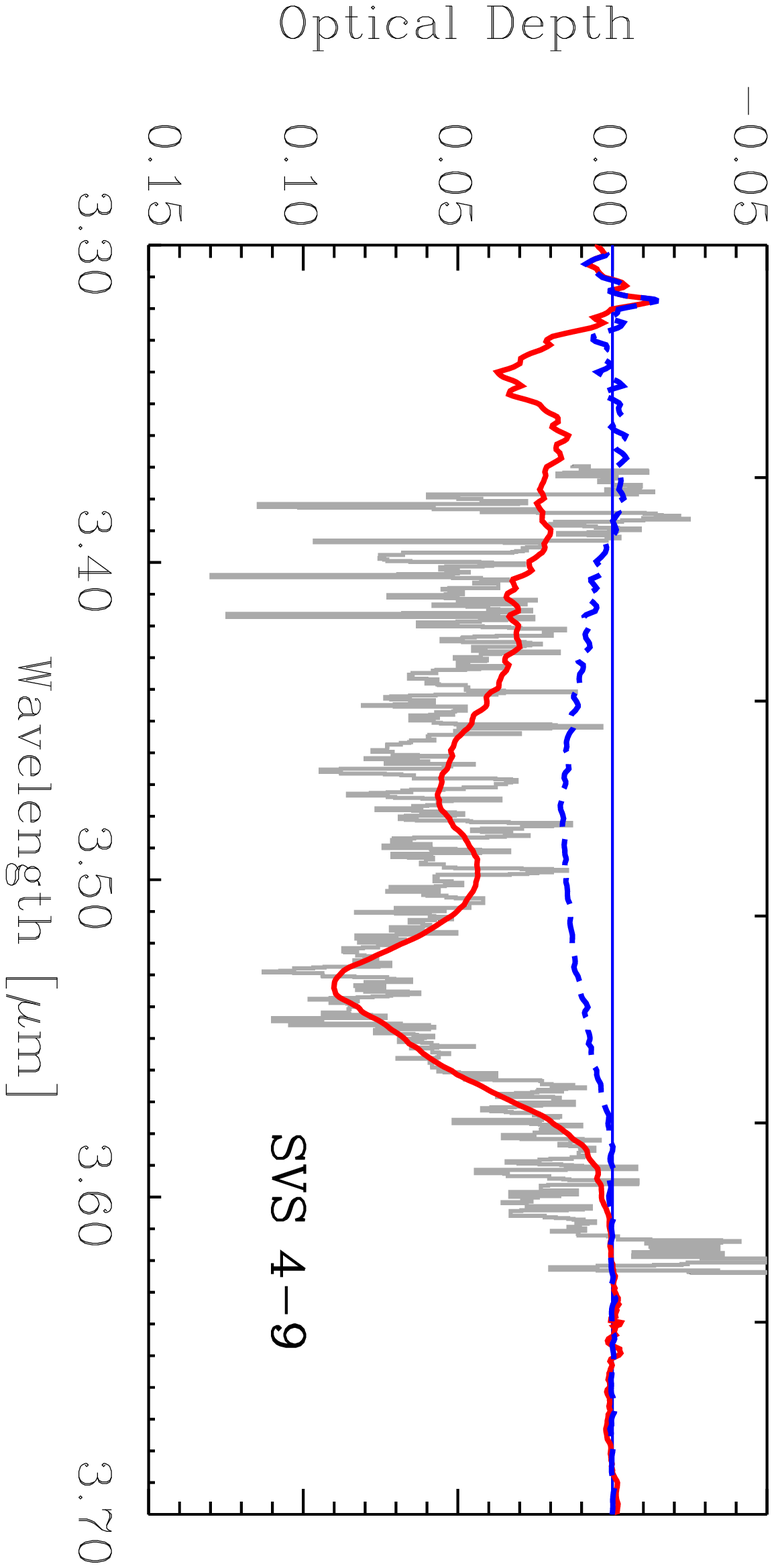}{3.0cm}{90}{30}{30}{100}{-25}
\caption{Top: Upper limit of the methanol 3.53\,$\mu$m band toward the
edge-on disk CRBR 2422.8-3423 ($N({\rm CH_3OH})/N({\rm H_2O})\la 0.06$).
Bottom: VLT-ISAAC detection of methanol toward a YSO (SVS 4-9) which probes dense
envelope material and an outflow rather than a disk ($N({\rm CH_3OH})/N({\rm H_2O}) = 0.25$).
The dashed curves show a template 3.47\,$\mu$m feature.
The solid curves show a sum of the 3.47\,$\mu$m template and a 
laboratory spectrum of an $\rm H_2O:CH_3OH:H_2CO$=1:1:1 mixture at 10 K.
}
\label{crbr}
\end{figure}

Here we present the results from a search for solid methanol toward these two sources. 
Methanol ($\rm CH_3OH$) is believed to be formed primarily in the solid 
state through successive hydrogenation of CO. 
The formation efficiency of this species is 
particularly interesting due to its significant role in the gas-phase formation of more complex organic
molecules. In Fig. 1, a VLT-ISAAC spectrum of the 3.53\,$\mu$m region of CRBR 2422.8-3423 is presented
showing a non-detection of solid methanol with an upper limit of 6\% with respect to 
water ice. For comparison a strong methanol band observed toward the low mass YSO
SVS 4-9 in the Serpens cloud core is shown below (Pontoppidan et al. 2003). 

\begin{figure}
\plotfiddle{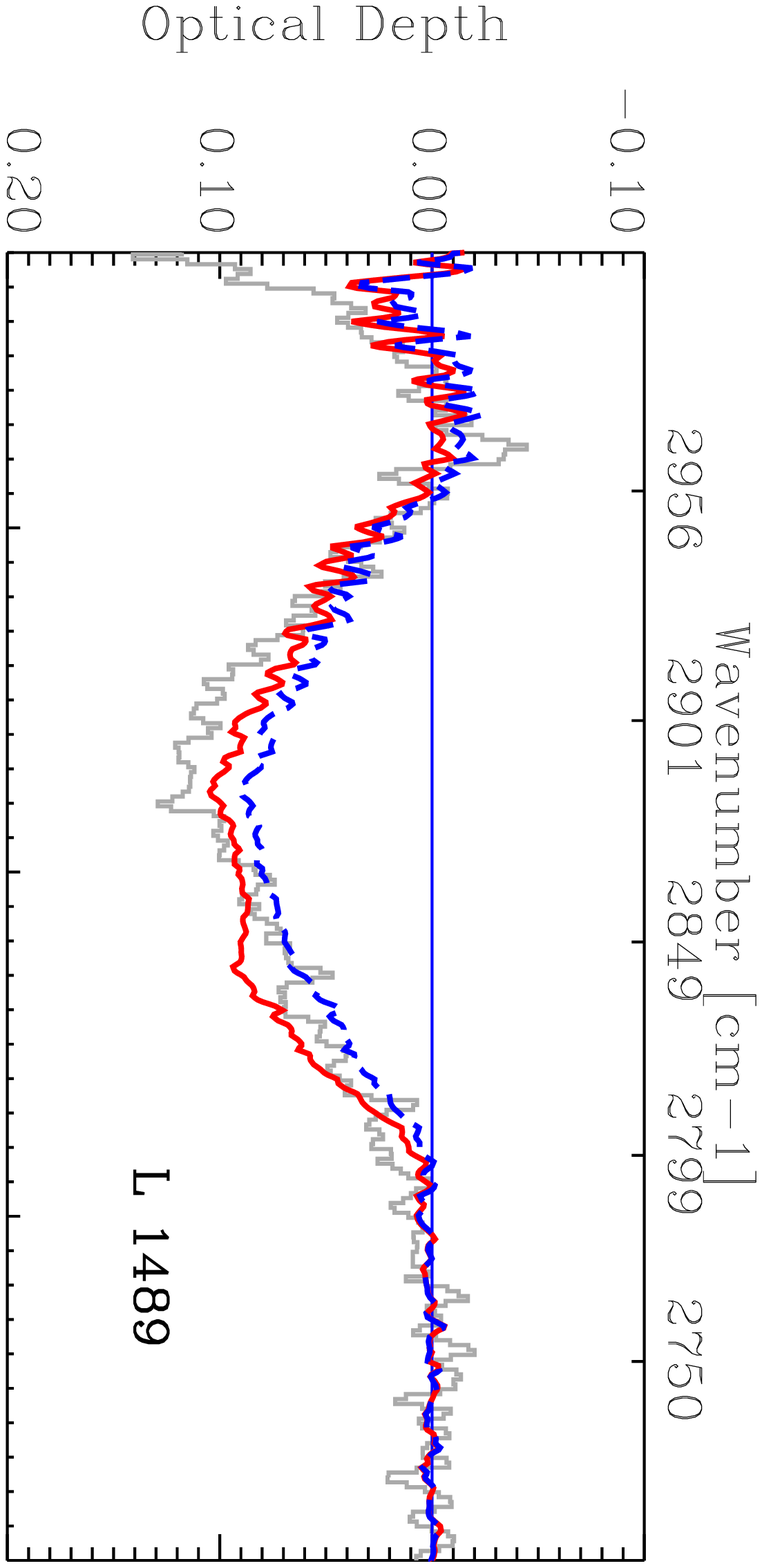}{3cm}{90}{30}{30}{100}{-40}
\plotfiddle{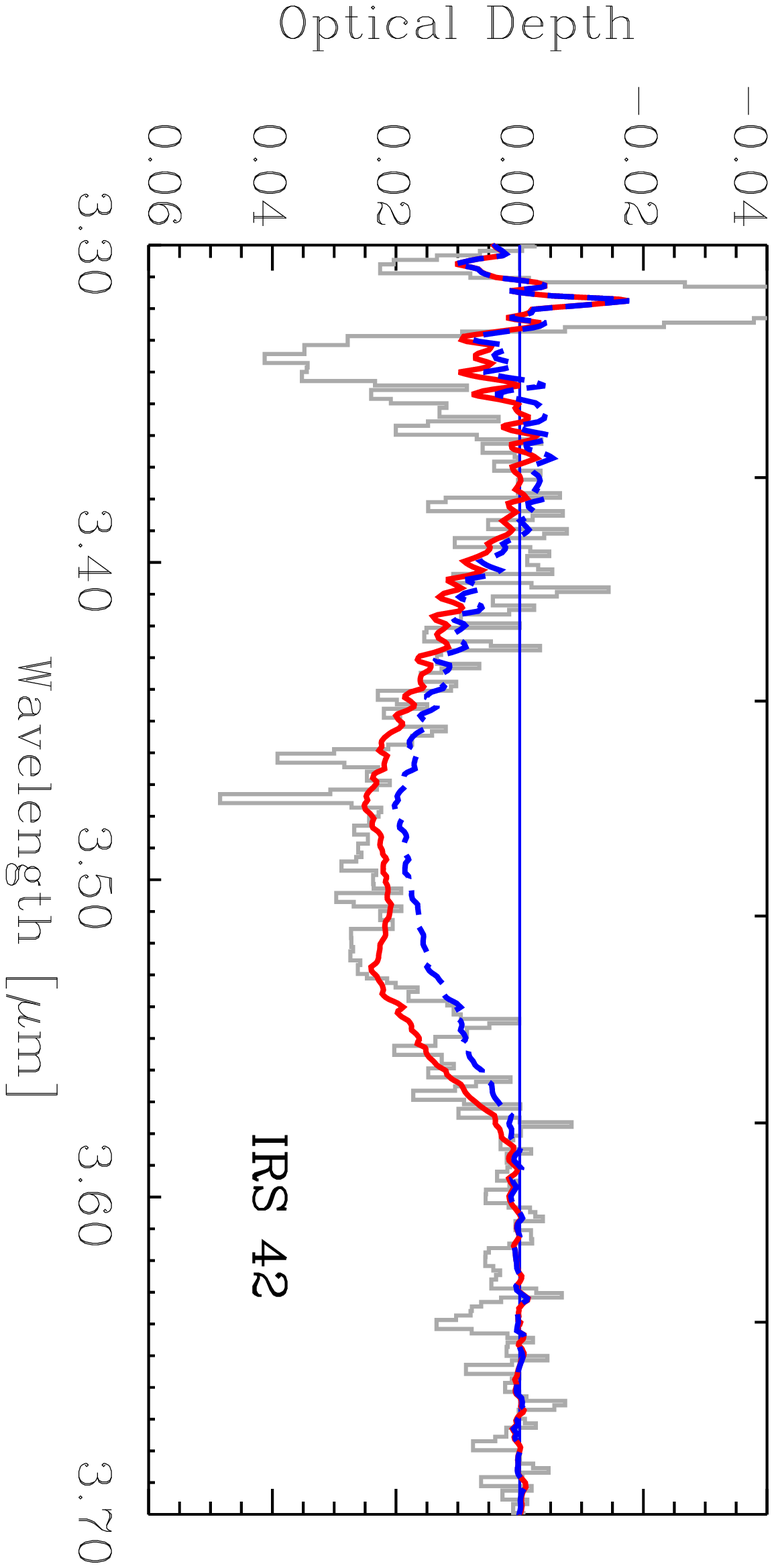}{3cm}{90}{30}{30}{100}{-25}
\caption{Top: The deep 3.47\,$\mu$m feature of the edge-on disk source L1489. The solid
curve corresponds to a methanol content of 5\%.
Bottom: A comparison with IRS 42 in the Ophiuchi cloud core thought to probe mostly quiescent
molecular cloud material. The solid curve corresponds to a methanol content of 4\%. See Fig. 1 for details curves.}
\label{crbr}
\end{figure}

In Fig. 2 the same spectral region is shown of L1489. The band is clearly dominated by the 
``3.47\,$\mu$m feature'', which is unrelated to the methanol band. The presence of this band makes is difficult 
to determine the abundance of solid methanol even if the spectrum is of high quality. The spectrum 
of another low mass source, IRS 42, is also shown as an example of a good upper limit of $\la 4$\%
methanol with respect to water ice. We find in general that about 10\% of our observed lines of sight
contain ice with 20\% or more methanol, while only upper limits of 2-10\% methanol are available for the rest.
This implies that the methanol content of specific grain mantles varies with at least an order of magnitude, 
depending on the environment. We have not found firm evidence for abundant methanol in the circumstellar
disk material observed so far, but our upper limits are still consistent with cometary abundances.
If methanol-rich ices from protostellar envelopes have been included in these disks, the methanol
must have been diluted or destroyed during the process of disk-formation.

\end{document}